\documentclass[
    ,final            
    ,sort&compress
  ]
  {aipproc}

\layoutstyle{8x11double}

\usepackage{natbib}

\newcommand{\Msun}{\mbox{$\mathrm{M}_{\odot}$~}}

\begin{document}

\title{ Binaries at Low Metallicity: \\ ranges for case A, B and C mass transfer}


\author{S.~E. de Mink}{
  address={Astronomical Institute Utrecht,
              Princetonplein 5, NL-3584 CC Utrecht, The Netherlands,
              S.E.deMink@astro.uu.nl}
}
\author{O.~R. Pols}{
address={Astronomical Institute Utrecht,
              Princetonplein 5, NL-3584 CC Utrecht, The Netherlands,
              S.E.deMink@astro.uu.nl}}

\author{S.-C. Yoon}{ 
address={Department of Astronomy \& Astrophysics, University of California, Santa Cruz, CA95064, USA}}

\begin{abstract}
The evolution of single stars at low metallicity has attracted a large
interest, while the effect of metallicity on binary evolution remains
still relatively unexplored. We study the effect of metallicity on
the number of binary systems that undergo different cases of mass
transfer. 

We find that binaries at low metallicity are more likely to start
transferring mass after the onset of central helium burning, often
referred to as case C mass transfer. In other words, the donor star in
a metal poor binary is more likely to have formed a massive CO core
before the onset of mass transfer.
At solar metallicity the range of initial binary separations that
result in case C evolution is very small for massive stars, because
they do not expand much after the ignition of helium and because mass
loss from the system by stellar winds causes the orbit to widen,
preventing the primary star to fill its Roche lobe.

This effect is likely to have important consequences for the
metallicity dependence of the formation rate of various objects
through binary evolution channels, such as long GRBs, double neutron
stars and double white dwarfs.

\end{abstract}

\maketitle


\section{Introduction}

During their life stars can expand to a radius which is up to 1000
times bigger than their initial radius. In close binary systems, they
start to transfer mass if their radius exceeds a critical radius, the
Roche lobe radius. It is usually the initially most massive star, the
primary star, which ``fills its Roche lobe'' first.

At solar metallicity binary evolution has been studied by various
groups, while their evolution at low metallicity is relatively
unexplored. Heavy elements, such as carbon, oxygen, nitrogen and iron,
are important contributors to the opacity of stellar material in metal
rich stars. Due to the lower opacity, metal poor stars are 
generally hotter and more compact. As it is the evolution of the
radius of a star in a binary which determines if and when mass
transfer occurs, we expect that the evolution of binaries in metal
poor environments is significantly different from binaries in the
solar neighborhood.

In this work we study the radius evolution of stars with different
masses in order to infer the frequency of various cases of binary
evolution as a function of metallicity. In a second contribution to
this proceedings \citep{acc07} we discuss the effect of metallicity on
the expansion of accreting main sequence stars and the potential
consequences for binary evolution.

\begin{figure}
  \includegraphics[bb=190 0 900 500, clip, width=\columnwidth]{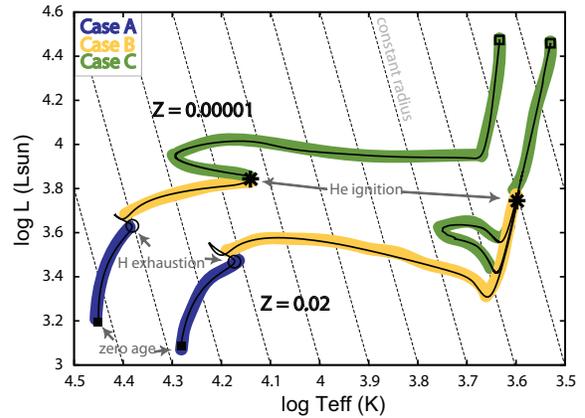}
  \caption{ Evolution track of a 6 \Msun star at solar (Z = 0.02)
    and low metallicity (Z =0.00001). 
    \label{hrd} }
\end{figure}

\begin{figure*}    
\includegraphics[ bb=50 10 680 480,
        width=\columnwidth]{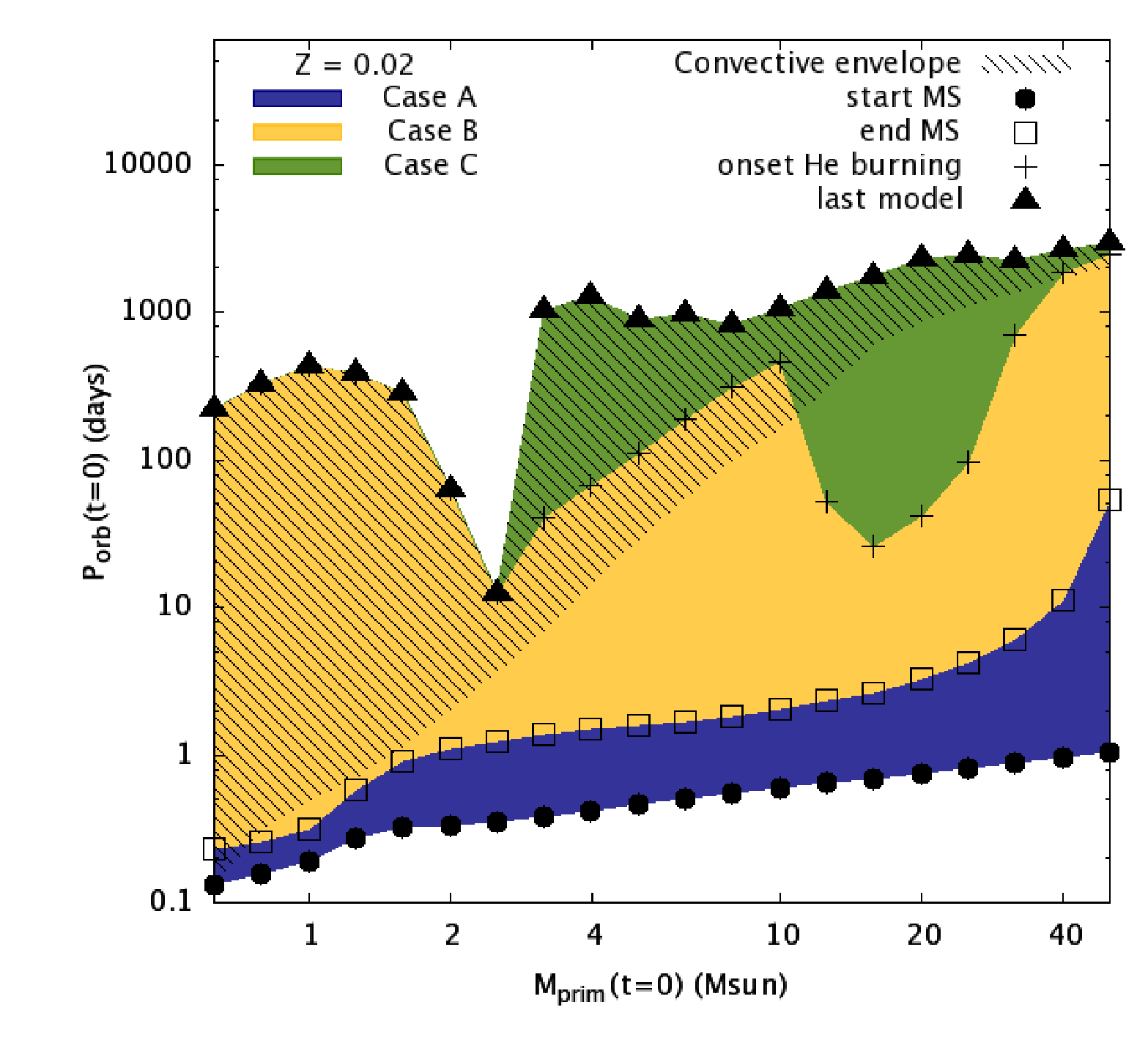}
\includegraphics[  bb=50 10 680 480,
        width=\columnwidth]{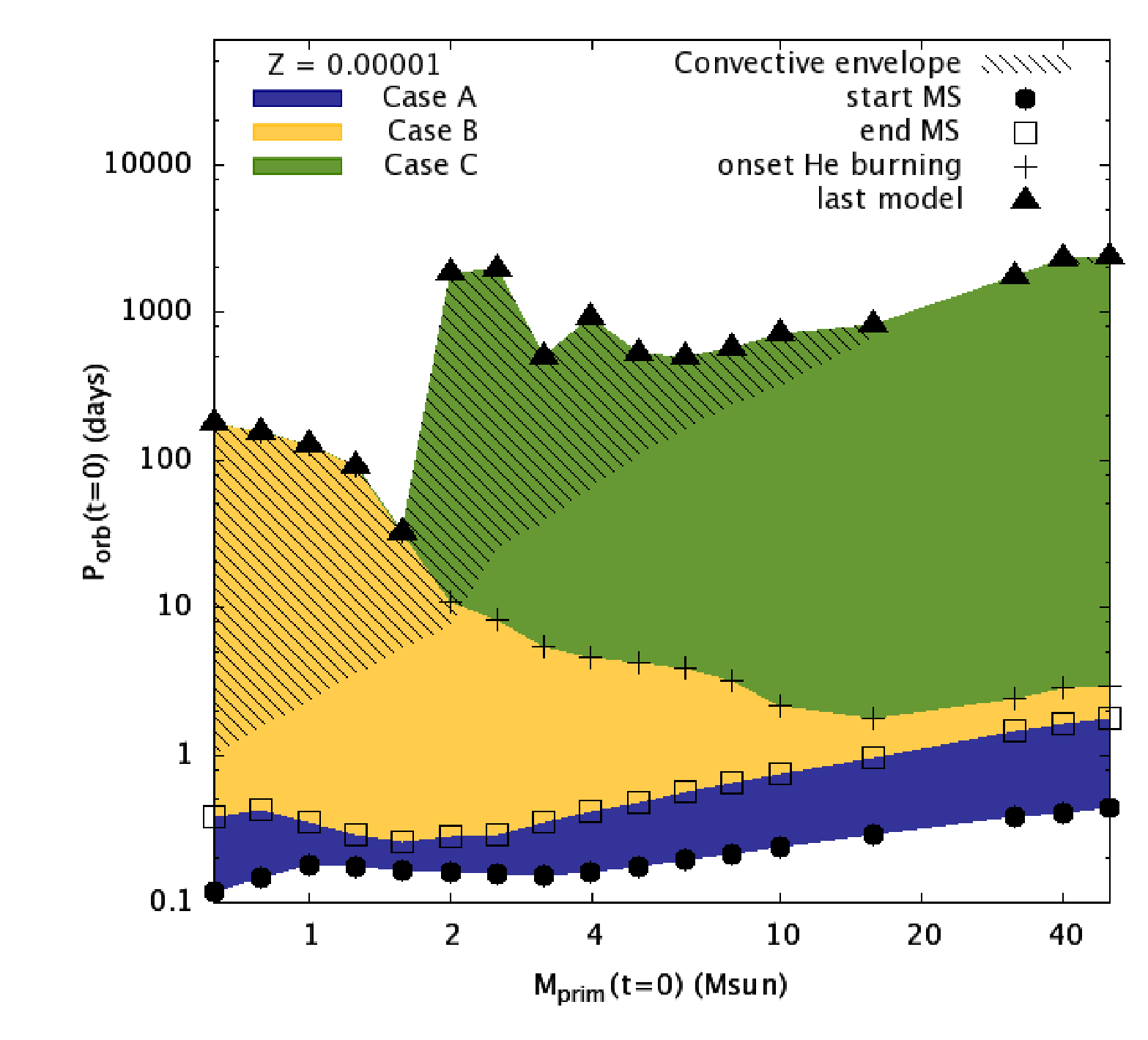}
\caption{ The initial orbital periods that lead to case A, B and C
mass transfer are given as function of the primary mass assuming an
initial mass ratio of 0.75 \citep[inspired by][]{Webbink79}. Widening
of the orbit due to angular momentum loss in the form of stellar winds
is taken into account. The hatching indicates approximately in which
cases the donor has a convective envelope.
        \label{cases} }
   \end{figure*}

\section{Evolution code}  
We use the fully implicit stellar evolution code \texttt{STARS},
originally developed by Peter Eggleton \citep{Eggleton71, Eggleton72,
Pols+ea95}, updated with the latest opacity tables
\citep{Eldridge+Tout04, OPAL96,Ferguson+Alexander05}.  For the
hydrogen and helium abundance we assume a linear relation with $Z$: $X
= 0.76 -3.0 Z$ and $Y = 0.24 + 2.0 Z$.  The abundances of the heavier
elements are assumed to scale to solar and meteoric abundance
\citep{Anders+Grevesse89} so that they are consistent with the opacity
tables.  A mixing length ratio $l/H_\mathrm{p}=2.0$ is
assumed. Convective overshooting is taken into account as in
\citet{Schroeder+ea97} with a free parameter $\delta_\mathrm{ov}=0.12$
calibrated against accurate stellar data from non-interacting binaries
\citep{Pols+ea97}. 

For the mass loss at solar metallicity we assume \citet{dejager88} and 
\cite{Reimers75}. We assume that the mass loss
scales with metallicity of the surface layers:
\[
\dot{M}(Z) = \dot{M}(Z_\odot) * \left(Z_{\rm surface} / Z_\odot
\right)^{0.8}
\]
\cite[inspired by][]{Vink+ea01}.
We want to emphasize that both the overshooting parameter and the
assumed mass loss, especially when the stars become red giants, are
important uncertainties in the model, which can have a large effect on
the radius evolution.

\section{Results}

Historically binary models have been classified as {\bf Case A} when
primary fills its Roche Lobe during its main sequence evolution, as
{\bf Case B} when it happens after H exhaustion, but before the
central ignition of Helium \cite{Kippenhahn+Weigert67} and as {\bf
Case C} when mass transfer starts after the ignition of Helium
\cite{Lauterborn70}.

Figure~\ref{hrd} shows the evolution track of a 6\Msun star in the
HR diagram at different metallicity. Throughout the main sequence the
low metallicity star is more compact and hotter. The circle indicates
exhaustion of H in the center (which we define as the moment that the
central abundance drops below $10^{-4}$). Ignition of He (when the
luminosity by He burning exceeds 5\% of the total luminosity) is
indicated with an asterisk. The low metallicity star ignites He at a
much smaller radius, while it is still in the blue part of the
diagram. The solar metallicity star first expands towards the giant
branch before it ignites Helium. This has implications for the
occurrence of case B and C evolution as a function of metallicity.

To determine the fraction of binaries that undergo case A, B and C
mass transfer, we calculated a set of stellar models with different
initial masses at $Z = 0.02$ and $Z=0.00001$ from which we determined
the maximum radius before H exhaustion and before He ignition. 
We can convert the radii to initial orbital periods, if we assume a
companion mass $0.75M_{\rm prim}$ and that the mass lost by stellar
winds from the primary take away the specific angular momentum of the
orbit of the primary, causing the system to widen, and if we neglect
mass loss from the secondary star
The result is shown in Figure \ref{cases}, where we plot which initial
orbital periods lead to case A, B and C mass transfer for different
masses of the primary. If we assume that binaries are formed with
initial orbital periods distributed uniformly in the log, then the
vertical width of the bands in the diagram directly represent the
fraction of binaries.

The blue band corresponds to the radius increase during main sequence
evolution, or to case A mass transfer. The yellow band corresponds to
case B evolution, which extends up to the maximum radius reached
before the onset of Helium burning. For stars more massive than 5\Msun
the range for case B becomes much smaller at low metallicity compared
to solar metallicity. The green band corresponds to case C evolution.
Beware that the upper boundary is given by the maximum radius during
our evolutionary calculations which is only a lower limit for the real
maximum radius that a star would reach.  The low mass stars, for
example, ignite helium in a degenerate core and we stopped our
evolutionary calculation. Case C evolution is likely to occur also for
low mass stars although it is not plotted in the diagram.

The hatching indicates in which systems the primary star will have a
convective envelope containing more than 15 \% of its mass. For these
systems mass transfer will occur on a dynamical timescale, causing the
accreting star to expand rapidly, probably resulting in a common
envelope situation.

The most striking difference is that the range for case C is much
larger at low metallicity.  This is partly caused by the fact that low
metallicity stars ignite He at a much smaller radius. A second
effect, which is most prominent at high masses, is that mass loss from
solar metallicity stars causes the orbit to widen, therefore decreasing
the initial periods leading to case C mass transfer. The range for
case C at solar metallicity becomes almost negligible for binaries
with primary masses above 30\Msun.

\section{Conclusion}
 
We find that at low metallicity the donor star in a binary is more
likely to have developed a massive CO core before the onset of mass
transfer. For binaries with primary masses above 30\Msun Case C mass
transfer may occur almost exclusively at low metallicity.

This effect potentially has important consequences for the metallicity
dependence of the formation rate of various objects through binary
evolution channels such as long Gamma Ray Burst progenitors \citep[see
discussion section of][]{wolf+podsiadlowski07} but also double neutron
stars and double white dwarfs. More work on binary evolution at low
metallicity is needed.

The precise fraction of Case C binaries as function of metallicity
depends on assumptions about the distribution of initial binary
parameters, the description of mixing and overshooting and the
(metallicity dependence) of the mass loss rate. We find that the
maximum radius can vary by more than a factor of 10 when we compare
calculations of the STARS code to calculations done with the code
described in \cite{Yoon+ea06} and when we vary the amount of
overshooting. In a forth coming paper we will present the range case
A, B and C mass transfer for a range of metallicities for different
assumptions for the uncertain parameters \citep{CASEABC08}.


\begin{theacknowledgments}
  We would like to thank Arend-Jan Poelarends, Matteo Cantiello, and
  Philip Podsiadlowski for interesting discussions and useful comments
  and suggestions.
\end{theacknowledgments}

\bibliographystyle{aipproc}   

\bibliography{deminkea_poster2}

\IfFileExists{\jobname.bbl}{}
 {\typeout{}
  \typeout{******************************************}
  \typeout{** Please run "bibtex \jobname" to optain}
  \typeout{** the bibliography and then re-run LaTeX}
  \typeout{** twice to fix the references!}
  \typeout{******************************************}
  \typeout{}
 }

\end{document}